\documentstyle[aps, preprint,prl]{revtex}  
\def\({\begin{equation}}
\def\){\end{equation}}
\begin{document}                
\title{Resistance of Josephson Junction Arrays at Low Temperatures}
\author{L. B. Ioffe, B. N. Narozhny}
\address{Department of Physics, Rutgers University, Piscataway, NJ 08855} 
\maketitle
\begin{abstract}  
We study motion of vortices in arrays of Josephson junctions at zero temperature where 
it is controlled by quantum tunneling from one plaquette to another. The tunneling 
process is characterized by a finite time and can be slow compared to the superconducting 
gap (so that $\tau \Delta >> 1$). The dissipation which accompanies this process 
arises from rare processes when a vortex excites a quasiparticle above 
the gap while tunneling through a single junction. We find that the dissipation is 
significant even in the case $\tau \Delta >> 1$, in particular it is not exponentially 
small in this parameter. We use the calculated energy dissipation for the single vortex
jump to estimate the physical resistance of the whole array.

\end{abstract}

\narrowtext
\section{Introduction}
In recent years dynamics of Josephson-junction arrays has attracted a lot of interest 
\cite{exp1,exp2,exp3,ecsc,mat,ann}. 
The Josephson-junction arrays (which are artificially fabricated networks of superconducting 
islands weakly coupled by tunnel junctions) became model systems for the study of quantum 
phase transitions, i.e. transitions occuring at $T\rightarrow 0$.  

The simplest physical picture of the phase transition in a 
two-dimensional short-ranged Josephson-junction array is the following.
The temperature is lower than the bulk transition temperature
of the islands, so that each individual island is superconducting and is characterized by a
phase of the superconducting order parameter. The absolute value of the order parameter, the
superconducting gap $\Delta$, is the largest energy scale in the problem.
The phase variable is conjugate to the
Cooper pair charge on the island. When the phase is well defined, the charge fluctuates and the 
array is superconducting. That happens in the limit where the Josephson energy $E_J$, associated 
with the Cooper pair tunneling, is much greater than the Coulomb energy $E_C$, which determines the 
electrostatic coupling between the islands that tends to localize the charge carriers.
In terms of vortices that means that in the limit $E_J \gg E_C$ the vortices form the 
Abrikosov lattice. In the opposite limit, $E_C \gg E_J$, 
the Coulomb blockade pins Cooper pairs to the islands,
so at low temperatures the array is insulating. Since in this phase the charge is fixed, the 
phase variable fluctuates and vortices form a superfluid.

Both phases were observed by preparing samples with different values of $E_C$ and 
$E_J$ \cite{exp1}. The insulating phase exhibits high values of resistance at finite temperatures,
which grow as $T\rightarrow 0$. The opposite behavior indicates the superconducting phase.
The transition can also be induced in the same sample by varying magnetic field.  
The field-induced transition can be experimentally observed in arrays \cite{exp2} and also 
in granular superconducting films \cite{exp3}. 

Conventional theoretical picture of the superconductor to insulator (S-I) transition suggested 
by M. P. A. Fisher \cite{mat} is based on duality between vortices and charges. In this picture
the transition point between the two phases is characterized by finite resistance, which is 
predicted to have universal value, proportional to the quantum resistance $R_q=h/4e^2$. 

Experimentally reported values\cite{exp1,exp2,exp3} of the transition point resistance, however, 
while being
of the same order as the predicted universal value, differ by as much as a factor of 5. 
Moreover, recent experiments \cite{exp} show that the superconducting and insulating phases are
separated by the wide metallic region, characterized by non-zero dissipation. In particular it 
was found that at low temperatures ($T<T_0=$100 mK) and in a noncommensurate magnetic field array 
resistance becomes temperature independent and remains finite down to the lowest temperatures 
accessible (10 mK). 

The metallic behavior of the arrays can not be described by the usual duality picture, since it 
ignores the presence of dissipation. Two issues have to be addressed. In terms of vortices,
a metal corresponds to a {\it normal liquid}, rather than the superfluid which
characterizes an insulator. Vortices, however, are interacting bosons and at low temperatures
tend to form the Bose condensate. Therefore the first question is how can the zero temperature 
normal liquid exist. The second question is what is the origin of dissipation at zero 
temperature.

In this paper we will focus on the second question. We consider vortex
motion at zero temperature where it is controlled by quantum tunneling 
of single vortices from one plaquette 
to another. It turns out that during the tunneling process a vortex can excite a quasiparticle state 
above the gap with the probability which is not exponentially small in the parameter $\tau \Delta >> 1$,
where $\Delta$ is the superconducting gap on the island and $\tau$ is the tunneling time. The 
relaxation of the excited quasiparticle then provides the dissipation in the system. 

In order to calculate the matrix element for quasiparticle excitation during vortex tunneling
we first solve a simpler quantum-mechanical problem.  
We consider a particle in a quasiclassical potential barrier which is also coupled to a single
harmonic oscillator. The probability to tunnel through the barrier is
given (in the simplest approximation) by the WKB approach. 
The initial state of the whole system (the particle and the oscillator) 
is that before particle tunneling the oscillator was in its ground state.
After the tunneling the oscillator could remain in its ground state
or it could be in one of its excited states; the latter case corresponds to
dissipation because for any non-zero coupling to environment the oscillator will eventually
relax to the ground state. Note that this relaxation can not affect 
the tunneling since it has already happened. For such a problem the dissipation is determined by
{\it conditional probability} of the oscillator excitation (given the fact of the tunneling)
which we calculate below in Section IV. 

The solution of this quantum mechanical problem can be applied to the case of vortex
tunneling. When vortex moves the phase on the islands changes. The time derivative of the phase
acts as an effective field acting on the quasiparticles and thus may result in quasiparticle
excitations. We note that the processes of excitation of different quasiparticle modes 
are independent, so the result for the total dissipation is given by the sum over all modes.

Besides the calculation of the matrix element, we have to make sure that in the process
of quasiparticle excitation the energy is conserved. In real arrays \cite{exp} the quasiparticle 
gap $\Delta$ is larger that both $E_C$ and $E_J$, so at zero temperature a single vortex does not have
enough energy to excite a quasiparticle. However in the vortex liquid dissipation does happen.
Experimental evidence \cite{efr} suggests that the vortex lattice melts easily due to 
frustration (since in the incommensurate magnetic field
the vortex lattice does not match the underlying array). 
That means that kinetic energy of vortices in the liquid is rather small, so that the liquid
retains the short range order. This viewpoint is supported by numerical simulations \cite{beda} of
2D melting, which show that below the melting point the vortices are mostly in large, ordered clusters.
When vortices move,
these clusters move as a whole and the extra momentum due to the interaction with quasiparticles 
in the island is transfered to the whole cluster. The energy of the cluster is much larger then the gap
in the quasiparticle spectrum, therefore the energy conservation is satisfied. 

The rest of the paper is organized as follows. In Section II we describe Josephson junction arrays 
in terms of phase variables. In the following Section we 
derive the form of the interaction between vortices and quasiparticles.
In Section IV we solve a quantum-mechanical problem of a particle in a barrier-like potential coupled
to a harmonic oscillator and in the next Section apply the results to the vortex tunneling. In
Sections VI we discuss the vortex lattice melting in the presence of frustration. In Section VII
we obtain the array resistance due to the dissipation found in Section V. The conclusions follow
in Section VIII.

\section{Josephson-junction arrays}
\label{sec:jos}

The microscopic theory of superconductivity in each individual island will be reviewed
in the next section; here we shall describe an array of small superconducting islands assuming that
the amplitude of the order parameter in each island
is constant and it is entirely controlled by a single phase variable, i.e. we ignore
its spatial dependence on the 
length scale of the size $a$ of the islands. This is true when the magnetic field does not penetrate
the bulk of the islands, which is guaranteed by the condition
that the flux through one island is less than the flux quantum
$Ha^2 < \phi_0$.

The array Hamiltonian consists in general of three parts. Time variations of the phase in each island
result
in voltage differences between the islands, which are electrostatically coupled to each other and to the
ground plane. That defines the first part of the Hamiltonian, the electrostatic energy as 
$(1/2)\sum\limits_{ij} \tilde{C}^{-1}_{ij} \dot{\phi}_{i}\dot{\phi}_{j}$, 
where $\tilde{C}_{ij}$ is proportional to a capacitance matrix 
$\tilde{C}^{-1}_{ij}=(2e)^2C^{-1}_{ij}$. 
In real experiment \cite{exp} the main contribution comes from the junction capacitance $C$.
Taking into account also the self-capacitance $C_0$ (the capacitance to the ground plane) we approximate
$C_{ij}$ by a matrix which only non-zero elements are diagonal $C_{ii} = C_0 + 4C$ and those corresponding 
to nearest neighbors in the array $C_{ij} = - C$. The junction capacitance defines the energy scale
$E_C = e^2/2C$, which is usually referred to as charging energy.

The second part of the array Hamiltonian is the Josephson coupling between the neighboring islands.
The coupling defines the other energy scale in the system $E_J$. The dissipation arises from the
coupling of the phase variable to some other degrees of freedom in the array. We will denote
that part of the Hamiltonian as $H_{int}$ and will derive its form in the next Section, where we
consider coupling of the time-dependent phase to the quasiparticles in the islands.
The Hamiltonian therefore is given by

\begin{equation}
H = {1\over2}\sum\limits_{ij} \tilde{C}^{-1}_{ij} \dot{\phi}_{i}\dot{\phi}_{j} - E_J
\sum\limits_{<ij>}\cos(\phi_i - \phi_j) + H_{int}.
\label{jham}
\end{equation}

At small Coulomb energy the ground state of Eq.\ (\ref{jham}) corresponds to the constant phase.
Excitations around that ground state are small spin-wave-like fluctuations and topological
defects, vortices, where the sum of all gauge-invariant phases around a vortex adds up to $2\pi$.
In the superconducting state, while the Coulomb energy is small, vortices appear in bound pairs 
(with anti-vortices). As the Coulomb energy increases, pairs unbound, resulting in the transition
to the insulating state of the array. In the field-tuned transition, $E_C$ is kept constant and
vortices are created by the magnetic field. In the superconducting state they form a lattice,
which melts at the transition point. Melting occurs at $E_C$ which is smaller then needed
to unbound the vortex-anti-vortex pairs. The density of vortices is controlled by magnetic field
and remains small even in the liquid phase. The flow of this vortex liquid 
results in finite resistivity. 

Vortices forming this liquid are distinguished by two important features. 
First, they do not have a normal core region, which would be the source
of dissipation in homogeneous superconductors. Second, since the Coulomb energy (which
is a measure for the kinetic energy of vortices) is small,
the vortex motion at low temperatures is due to quantum mechanical tunneling through the
cosine potential. 

The tunneling rate can be determined by calculating the instanton
action corresponding to a vortex moving from one site to the neighboring site. The instanton action
was determined by several authors \cite{kor,gei,faz} each for slightly different models 
without dissipation, which also differ from Eq.\ (\ref{jham}) by another form
of capacitance matrix; note, that theoretical calculation \cite{kor} involves also approximation of
the cosine Josephson interaction by a piecewise parabolic potential, i.e. the Villain's approximation. 
The results are similar, the 
instanton action is $S_{inst}=\alpha \sqrt{E_J/E_C}$, where the number $\alpha$ is of the order of
unity and depends on a particular model. To determine the resistivity in the array we need to take
into account also the dissipation (described by the $H_{int}$ term in Eq.\ (\ref{jham})), which is
the main subject of this paper. 

For better comparison with experiments we need to determine the constant $\alpha$ in the realistic
model with experimental values of coupling constants.
We have repeated the direct numerical instanton action calculation 
for the Hamiltonian Eq.\ (\ref{jham}) without the dissipation term $H_{int}$. 
First we find the phase configuration in the array corresponding to one vortex in a particular 
(arbitrary, but known) plaquette. To do that we set the magnetic field through the array so that the
total flux is exactly equal to one flux quantum and then minimize the energy Eq.\ (\ref{jham}).
Tunneling corresponds to changing the phase configuration to the one with the vortex in the
neighboring plaquette. In terms of phases the vortex tunneling
can be described as tunneling of each individual phase in the vortex configuration from it's value
corresponding to the original position of the vortex to the value corresponding to the final position
of the vortex. 
We set these two vortex configurations as the boundary conditions for time
evolution of individual phases in the array and minimize the action, 
corresponding to the Hamiltonian Eq.\ (\ref{jham}),
taking the phases to be functions of imaginary time. 

For array sizes 6$\times$6,
8$\times$8 and 10$\times$10 we determined the value of the coefficient $\alpha=0.7$ (for $C_0 = 0$). 
The phases that change the most during tunneling are the ones in the plaquette with the vortex. 
Therefore even with relatively small array sizes the calculation gives the answer that does not 
change with increasing the size.

The tunneling rate $\Gamma_0\sim\exp(-S_{inst})$ is then given by

\begin{equation}
\Gamma_0 \sim {1\over{\hbar}} { \; } {{\sqrt{E_JE_C}}} \exp(-0.7 \sqrt{E_J/E_C})
\label{trate}
\end{equation}

\noindent
and provides a measure for the vortex mass.

\section{Superconductivity in a single island}

In this section we briefly derive the $H_{int}$ part of the phase Hamiltonian Eq.\ (\ref{jham})
which couples phase fluctuations to quasiparticles. In this derivation we follow
the standard microscopic description of superconductivity based on the
BCS Hamiltonian \cite{bcs,sch}.

We start with the BCS Hamiltonian with an effective local, attractive interaction

\begin{equation}
H_{BCS} = - \int d^3x \psi^\dagger_\sigma(x) {\nabla^2\over{2m_e}}  \psi_\sigma(x)
-{g_{0}\over2} \int d^3x \psi^\dagger_\sigma(x) \psi^\dagger_{-\sigma}(x) \psi_{-\sigma}(x) \psi_\sigma(x).
\end{equation}

\noindent
A summation over spins is implied. The order parameter $\Delta(x, \tau)$ is introduced by means of
the Hubbard-Stratonovich transformation. The grand canonical partition function 
${Z_G=Tr_\psi \{\exp[-\beta(H-\mu N)] \} }$ becomes 

\begin{equation}
Z_G = Tr_\psi \left \{ \int {\cal D}[\Delta, \Delta^*] \; T\exp\left( -\int\limits_0^\beta d\tau
H_{eff}(\tau)\right) \right \},
\label{part}
\end{equation}

\noindent
where the effective Hamiltonian is given by

\begin{eqnarray}
H_{eff}(\tau) \; &&= \int d^3x \LARGE\{\psi^\dagger_\sigma(x) \left(-{\nabla^2\over{2m_e}}  - \mu\right)
\psi_\sigma(x)  \\
&&
\nonumber\\
&& - \Delta^*(x, \tau) \psi_\uparrow(x)\psi_\downarrow(x) - 
\Delta(x, \tau) \psi^*_\downarrow(x)\psi^*_\uparrow(x) + {1\over g_{0}} |\Delta(x, \tau)|^2 \LARGE\}.
\end{eqnarray}

\noindent
We can compactify our notations by introducing Nambu spinor \cite{nam}

\begin{eqnarray}
\hat{\psi} = 
\pmatrix
{
{  \psi_\downarrow}\cr
{  \psi^*_\uparrow}\cr
}.
\end{eqnarray}

\noindent
The effective Hamiltonian becomes

\begin{eqnarray}
H_{eff}(\tau) = \int d^3x \left[ \hat{\psi}^\dagger \left( {\it K} \hat{\tau}_3 - \hat{\Delta}
\right)\hat{\psi} + {1\over g_{0}} |\Delta(x, \tau)|^2 \right],
\end{eqnarray}

\noindent
where ${\it K} = - \displaystyle{ \nabla^2 \over{2m_e}} - \mu$ is the kinetic operator and 
the order parameter is described by the matrix 
$ \hat{\Delta} = |\Delta| \displaystyle{e^{\displaystyle{-i \phi \hat{{\tau}_3}}}} \hat{\tau}_1$, 
$|\Delta|$ and $\phi$ are the absolute value and the phase of the order parameter. In our approximation
$|\Delta|$ is constant and $\phi(\tau)$ is a function of time.

The Hamiltonian is now quadratic in fermion fields and we can formally perform the trace 
over the fermion variables in 
Eq.\ (\ref{part}). The partition function becomes the integral over the order parameter

\begin{equation}
Z_G = \int {\cal D}[\Delta, \Delta^*] \; \exp\left( - S[\Delta] \right),
\end{equation}

\noindent
where the action is 

\begin{equation}
S[\Delta] = - Tr \ln \hat{G}^{-1} + \int\limits_0^\beta d\tau {1\over g_{0}} |\Delta(\tau)|^2.
\label{aaa}
\end{equation}

\noindent
Here $\hat{G}$ is a 2$\times$2 matrix Green's function in the particle-hole space \cite{nam}
typical for superconductivity which inverse is given by

\begin{equation}
\hat{G}^{-1} (x, \tau; x', \tau') = 
\left\{ -{\partial\over{\partial\tau}} - {\it K} \hat{\tau}_3 + \hat{\Delta} \right\}
\delta(x - x^\prime)\delta(\tau - \tau^\prime).
\label{grin}
\end{equation}

\noindent
The action Eq.\ (\ref{aaa}) is a standard BCS action written in the form convenient for the 
following. It does not contain any dissipation as
yet, therefore there are no non-local terms discussed in Ref. \cite{sch}. The
dissipation appears after an additional assumption about time
dependence of the phase variable, namely that while on average it
changes slowly, this change occurs with rare but large enough jumps,
due to the lattice structure of the array. Thus in the following we shall not assume that 
$\phi(t)$ is a smooth function of time;
such assumption would eliminate all dissipation sources in this problem.

The dependence of the Green's function (and therefore the action) on the phase of
the order parameter can be displayed through the gauge transformation

\begin{equation}
{\cal G}^{-1} (x, \tau; x', \tau') = e^{\displaystyle{-i \phi \hat{{\tau}_3}/2}}
\hat{G}^{-1} (x, \tau; x', \tau') e^{\displaystyle{i \phi \hat{{\tau}_3}/2}},
\end{equation}

\noindent
where ${\cal G}^{-1}$ is obtained from Eq.\ (\ref{grin}) by the replacement 
$\displaystyle{{\partial\over{\partial\tau}}
\rightarrow {\partial\over{\partial\tau}}-{i\over2}{{\partial\phi}\over{\partial\tau}}\hat{\tau}_3}$.
This transformation shows that a constant $\phi$ contributes nothing to the to the action Eq.\ (\ref{grin}).

The fermion contribution to the action can be represented by the path integral over Grassman variables

\begin{equation}
\exp(Tr \ln {\cal G}^{-1}) = \int {\cal D}[\psi_\sigma] \exp( - S_\psi),
\end{equation}

\noindent
where the fermion action 

\begin{equation}
S_\psi = \int d\tau d\tau' \int d^3x d^3x' \;  \hat{\psi}^\dagger {\cal G}^{-1} (x, \tau; x', \tau') \hat{\psi}
\end{equation}

\noindent
is explicitly given by

\begin{equation}
S_\psi = \int d\tau d^3x \;  \hat{\psi}^\dagger 
\left[ - ({\partial\over{\partial\tau}}-{i\over2}{{\partial\phi}\over{\partial\tau}}\hat{\tau}_3)
 - {\it K} \hat{\tau}_3 + |\Delta|\hat{\tau}_1 \right] \hat{\psi}. 
\end{equation}

At zero temperature we can write the real time action as 

\begin{equation}
S_\psi = i \int dt d^3x \;  \hat{\psi}^\dagger 
\left[ - i{\partial\over{\partial t}}-{1\over2}{{\partial\phi}\over{\partial t}}\hat{\tau}_3
 - {\it K} \hat{\tau}_3 + |\Delta|\hat{\tau}_1 \right] \hat{\psi}. 
\end{equation}

\noindent
The corresponding Hamiltonian in momentum space is given by the 2$\times$2 matrix

\begin{eqnarray}
H_\psi = 
\pmatrix
{
{-\epsilon_k-\varphi}&{|\Delta|}\cr
{|\Delta|}&{\epsilon_k+\varphi}\cr
}.
\end{eqnarray}

\noindent
where $\varphi = \displaystyle{{1\over2}{{\partial\phi}\over{\partial t}}}$. In the BCS theory $\varphi=0$ and
the Hamiltonian can be diagonalised by the Bogolyubov transformation, which is just a rotation of
the fermion variables. When  $\varphi \neq 0$ we still can perform the rotation, but the 
resulting action will no longer be diagonal due to the time dependence of $\varphi$. The rotation
matrix, which diagonalises the Hamiltonian at each moment of time is 

\begin{eqnarray}
{\cal R} = {1\over\displaystyle{\sqrt{2\lambda_k(\lambda_k+\epsilon_k+\varphi)}}}
\pmatrix
{
{\lambda_k+\epsilon_k+\varphi}&{-|\Delta|}\cr
{|\Delta|}&{\lambda_k+\epsilon_k+\varphi}\cr
}.
\end{eqnarray}

\noindent
After the rotation the Hamiltonian becomes diagonal

\begin{eqnarray}
\hat{{\cal H}} = 
\pmatrix
{
{-\lambda_k}&{0}\cr
{0}&{\lambda_k}\cr
},
\end{eqnarray}

\noindent
with the eigenvalue $\lambda_k = \displaystyle{\sqrt{(\epsilon_k+\varphi)^2 + |\Delta|^2}}$. The action
becomes

\ 

\begin{equation}
S_\gamma = i \int dt d^3k \;  \hat{\gamma}^\dagger(k, t) 
\left[ - i{\partial\over{\partial t}} + \hat{{\cal H}} 
 - {i\over\displaystyle{2\lambda_k(\lambda_k+\epsilon_k+\varphi)}}  
\pmatrix
{
{\epsilon_k+\varphi}&{-|\Delta|}\cr
{|\Delta|}&{\epsilon_k+\varphi}\cr
}
{{\partial\varphi}\over{\partial t}}\right] \hat{\gamma}(k, t), 
\label{action}
\end{equation}

\ 

\noindent
where $\hat{\gamma}(k, t)$ are the variables in the rotated basis (which for $\varphi=0$ correspond 
to Cooper pairs). 

The additional term in Eq.\ (\ref{action}) appeared due to the time dependence of $\varphi$. 
The diagonal part is small compared to the eigenvalues $\lambda$ and can be ignored. 
The non-diagonal part, however, describes a new process : a quasiparticle excitations (the Cooper
pairs correspond to diagonal part of Eq.\ (\ref{action}) ). This is the interaction term which is
responsible for dissipation. Upon integrating out the fields $\hat{\gamma}(k, t)$ it becomes 
the interaction part of the action, corresponding to
the $H_{int}$ part of the array Hamiltonian Eq.\ (\ref{jham}).

\begin{equation}
S_{int} = i \int dt d^3k \;  \hat{\gamma}^\dagger(k, t) 
{1\over\displaystyle{2(\lambda_k+\epsilon_k)}}  
\pmatrix
{
{0}&{-|\Delta|}\cr
{|\Delta|}&{0}\cr
}
{{\partial\phi}\over{\partial t}} \hat{\gamma}(k, t), 
\label{intact}
\end{equation}

\noindent
Here we have integrated the interaction term by parts, in order to express the result in
terms of the phase fluctuations $\phi$. This brings the extra factor $2\lambda_k$ from the
time dependence of $\hat{\gamma}(k, t)$. Also we neglected $\varphi$ in the prefactor which
forms the coupling constant 
$g_k\approx\displaystyle{{|\Delta|}\over{2(\lambda_k+\epsilon_k)}}$ because
we consider only adiabatically slow motion.

The phase now can be treated as independent variable, describing a ``particle''
in the periodic potential and coupled to the quasiparticles in the island 
through the action Eq.\ (\ref{intact}). Note that the action Eq.\ (\ref{intact}) is diagonal 
in momentum $k$, so for each $k$ the quasiparticle action is that of a two-level system. 
Since the probability to excite a quasiparticle is small, the phase fluctuations excite 
only one two-level system at a time, so these excitation processes are independent and
the total probability can be found as a sum of probabilities to excite each individual
two-level system. Therefore we can consider the quantum mechanical
problem of a particle coupled to the two-level system and then integrate the results
over $k$.

Furthermore, dissipation resulting from exciting a two-level system is
not different from the one of an oscillator because excitations of the latter to higher 
levels can be neglected. The latter problem has a slightly broader application. Note, however,
the important difference between this problem and the Caldeira-Legget \cite{cal} model.
Here the motion of the particle is coupled to a single oscillator with large level spacing, which 
corresponds to the large quasiparticle gap, whereas the 
Caldeira-Legget \cite{cal} model is a system 
coupled to a large number of small oscillators, so it's not difficult to excite each
one individually and interesting physics arises from exciting a large number of 
them simultaneously.

Thus we reduced our problem of calculating the probability to excite a quasiparticle
in an island to a problem of exciting a harmonic oscillator during tunneling. In the next 
section we consider this simpler problem and then in the following section we apply the
obtained results to the case of coupling to a two-level system and then sum the probability
over momenta $k$ to obtain the final probability to excite a quasiparticle.

\section{Simple model - particle coupled to harmonic oscillator}

In this section we consider the quantum mechanical problem of a particle coupled 
to a harmonic oscillator and
tunneling through some barrier. The Hamiltonian is

\begin{equation}
H = {\hat{p}^2\over{2m}} + V(x) + {\hat{P}^2\over{2M}} + {1\over2}M\omega_0^2Q^2 + g\hat{p}Q,
\label{osh}
\end{equation}

\noindent
where $\hat{p}$ and $\hat{P}$ are momentum operators of the particle and the oscillator, $m$ and $M$ are
their respective masses, $V(x)$ is the potential which we assume has a form of a barrier 
and $g$ is the
coupling constant. Here we chose the coupling coupling (using the momentum operator $\hat{p}$ rather then the 
position operator) which has the same form as the one in
the action Eq.\ (\ref{intact}). We need to obtain the 
probability to find the oscillator in its first excited state after the particle have tunneled through
the barrier if before the tunneling the oscillator was in the ground state. The coupling $g$ is taken
to be small enough so that the oscillator states are unchanged.

We look for the wave function of the system in the form 

\begin{equation}
\Phi = \sum\limits_n \Psi_n(x) \; |n\rangle,
\end{equation}

\noindent
where $|n\rangle$ denotes oscillator wave functions corresponding to $n$-th energy level. 

Neglecting
the coupling completely we should have

\begin{equation}
\Phi_0 = \Psi_0(x) \; |0\rangle,
\end{equation}

\noindent
which means that the oscillator is in the ground state. The particle wave function under the barrier
is given in the WKB approximation by

\begin{equation}
\Psi_0(x) \approx  u_0(x) \exp \left( - \int\limits_{x_a}^{x} \sqrt{2m(V-E)} dx \right).
\label{wkb}
\end{equation}

In the first order in $g$ we have for the wave function $\Phi_1=\Psi_0(x)|0\rangle + \Psi_1(x)|1\rangle$.
The Schr\"odinger equation for the correction $\Psi_1(x)$ is

\begin{equation}
\left[ -{1\over{2m}} {\partial^2\over{\partial x^2}} + V(x) - E + \omega_0 \right] \Psi_1(x) -
 i g Q_{10}{\partial\over{\partial x}} \Psi_0(x) = 0,
\end{equation}

\noindent
where $Q_{10} = \langle 1 | Q |0 \rangle = 1/\sqrt{2M\omega_0^2}$ is the oscillator matrix element. 
The oscillator ground state energy $\omega_0/2$ is included in the definition of $E$. It is convenient
to express the solution in the form $\Psi_1(x) = u_1(x) \Psi_0(x)$ with the boundary condition
$u_1(x_a) = 0$, noting that the oscillator was in the ground state prior to tunneling. The equation
becomes

\begin{equation}
\displaystyle{\Psi_0\over{2\Psi_0^\prime}} u_1^{\prime\prime} + u_1^{\prime} - m \omega_0 
\displaystyle{\Psi_0\over{\Psi_0^\prime}} u_1 + i m g Q_{10} = 0.
\label{equ}
\end{equation}

Compare now first and second derivative terms. The typical particle energy is 
${\Omega = \displaystyle{\sqrt{V\over{mL^2}}}}$, where $L\sim(x_b-x_a)$.
The typical time $\tau \sim 1/\Omega$. The ratio $\displaystyle{\Psi_0\over{\Psi_0^\prime}}$ can be 
estimated using Eq.\ (\ref{wkb}) as $\displaystyle{1\over\sqrt{mV}}$.
Therefore the  second derivate term can be estimated as 
$\displaystyle{\Psi_0\over{2\Psi_0^\prime}} u_1^{\prime\prime} \sim u_1\displaystyle{1\over{L^2\sqrt{mV}}}$.
The first derivative term is simply $u_1^{\prime} \sim u_1/L$, so that their ratio is 
$\displaystyle{\Psi_0\over{2\Psi_0^\prime}} u_1^{\prime\prime} / u_1^{\prime} \sim 
\displaystyle{1\over{L\sqrt{mV}}} \ll 1$. Therefore we can drop the second derivative term in
Eq.\ (\ref{equ}). Solving the remaining first order equation we get for the function $u_1$ right after 
the tunneling

\begin{equation}
u_1(x_b) = - i m g Q_{10} \int\limits_{x_a}^{x_b}
\left[ \exp \left( - \int\limits_{x}^{x_b} {{m\omega_0}\over{\sqrt{2m(V-E)}}} ds \right) \right] dx.
\label{soleq}
\end{equation}

Consider now two limiting cases. When $\omega_0 \ll \Omega$ (fast transition), the integrand in the 
exponential in Eq.\ (\ref{soleq}) is small, therefore $u_1(x_b) \approx - i m g Q_{10} L$ so there is
no additional suppression other than the smallness of $g$. In the opposite limit $\omega_0 \gg \Omega$
we can evaluate the integral in the exponential noting that the main contribution comes from the 
region near the ending point $x_b$. We can then expand the integrand around $x_b$ to get
${\displaystyle{{{m\omega_0}\over{\sqrt{2m(V-E)}}}} \sim 
\displaystyle{{{2m\omega_0}\over{\sqrt{2mV'}}}\sqrt{x_b-x}}}$, where 
$V' = \displaystyle{{dV}\over{dx}}\Biggm{|}_{x=x_b}$. Evaluating the integral is now straightforward
and we get 

\begin{equation}
u_1(x_b) = - i m g Q_{10} L {V'\over{m\omega_0L}} 
\sim - i m g Q_{10} L \left( {\Omega\over\omega_0} \right)^2,
\label{sol}
\end{equation}

\noindent
which means that in the limit of slow tunneling $u_1(x_b)$ is indeed small in $\Omega/\omega_0$ but only
as a power law.

The probability of exciting the oscillator is proportional to the square of $u_1(x_b)$ and is given by

\begin{equation}
{\cal P} \approx g^2 m^2 L^2 Q_{10}^2 \left( {\Omega\over\omega_0} \right)^4.
\label{pro}
\end{equation}

\noindent
The average energy ${\cal W}$ dissipated in one jump
is equal to the energy needed to excite the 
oscillator ($\omega_0$) times the transition probability Eq.\ (\ref{pro})

\begin{equation}
{\cal W} \approx (g m L Q_{10})^2 \; \omega_0 \left( {\Omega\over\omega_0} \right)^4.
\label{diss}
\end{equation}

\section{Tunneling of particle coupled to quasiparticle system}

We now return to the full problem, formulated in Section ~\ref{sec:jos}, namely to
the action Eq.\ (\ref{intact}). As we have already mentioned, for each value of $k$ we can treat the 
quasiparticle system as a two-level system and then sum over all possible $k$. Again we treat the 
phase variable on the island as a coordinate of a ``particle'', which tunnels through some barrier.
The result for this case is exactly the same as for the case of the oscillator since we have
neglected excitations of higher levels. In the case of a two-level system there are no higher
levels at all and the result Eq.\ (\ref{diss}) is the full answer, in which we have to
substitute the corresponding matrix element for $Q_{10}$ and the value of the gap for $\omega_0$.

Therefore in order to apply our results to the case of the phase coupled to the quasiparticle system 
Eq.\ (\ref{intact}) we only need to rewrite it in a Hamiltonian form equivalent to Eq.\ (\ref{osh}).
In the Hamiltonian formalism the derivative $\displaystyle{{\partial\phi}\over{\partial t}}$
in Eq.\ (\ref{intact}) is replaced by the
momentum operator $\hat{p}/m$ ($m$ is the ``particle'' mass) leading to the effective interaction
constant

\begin{equation}
g_k = {1\over m}{{|\Delta|}\over{2(\lambda_k+\epsilon_k)}}.
\end{equation}

\noindent
The energy $\Omega$ is now the typical frequency of the phase variation and is equal to 
the inverse tunneling time, $\Omega\approx\sqrt{E_JE_C}$.

The probability to excite the quasiparticle at each $k$ then follows from Eq.\ (\ref{pro}). 
The oscillator frequency $\omega_0$ is now exchanged for the quasiparticle gap, 
which is $2\lambda_k$ at the same $k$. The matrix element
corresponding to $Q_{10}$ is just 1. The probability then is 

\begin{equation}
{\cal P}_k \approx (g_k m 2\pi)^2 \left( {\Omega\over {2\lambda_k}} \right)^4.
\label{kpro}
\end{equation}

\noindent
Here we used the factor of $2\pi$ for the effective length $L$, which is by how much the phase can
be changed. The dissipation contribution for each $k$ follows by multiplying the 
probability be the energy gap

\begin{equation}
{\cal W}_k \approx (g_k m 2\pi)^2 \; 2\lambda_k \left( {\Omega\over{2\lambda_k}} \right)^4.
\end{equation}

\noindent
Integrating this expression over momentum we get

\begin{equation}
{\cal W} \approx  \pi^5 \Omega \left( {\Omega\over{|\Delta|}} \right)^3 {\cal N},
\label{resultd}
\end{equation}

\noindent
where ${\cal N}$ is the number of particles on the island. This defines
the energy transfered to the quasiparticle system during tunneling and therefore 
the dissipation in a single vortex jump between two neighboring plaquettes.

\section{Vortex liquid}

At low magnetic fields vortices form a lattice that melts at higher fields. Because melting
is due to the competition between kinetic and interaction energies, 
it happens when the two are parametrically equal. However in 2D the liquid retains 
short-range order and the interaction energy loss in melting is numerically much smaller as 
described by a small Lindemann number. In arrays the vortex lattice is frustrated by the
incommensurability with the underlying array structure. This effect reduces the 
ratio of kinetic energy over interaction 
energy at melting even further.

In the absence of a microscopic theory of melting we use the
phenomenological Lindemann criterion, which describes the melting in terms of elastic 
constants of the vortex lattice. In the array system these constants are renormalized by
frustration. To estimate this effect we analyze the experiment on thermal melting \cite{efr},
in which the effect of incommensurability on the transition temperature was studied in
detail. We emphasize that these measurements were performed on array systems, which are
different from the ones discussed throughout this paper. Here we use these experiments  
to obtain estimates of the renormalization of the elastic constants of the vortex lattice and 
then use this renormalization to describe quantum melting. Our observations however are general and
therefore are applicable to the quantum systems of interest \cite{exp}.
 
We need to estimate how frustration renormalizes the elastic constants. Therefore we
estimate the interaction energy in the experimental system \cite{efr}, which translates
into unrenormalised values of elastic constants, leading to an estimate of unrenormalised
melting temperature $T_{m0}$. Comparing $T_{m0}$ with the experimentally observed $T_{m}$
we find the frustration factor.

To estimate the interaction energy we relate the superfluid density $\rho_s$ to the observed
magnetization $M$. Energy and current in the Josephson junction in magnetic field can be 
written as

\begin{equation}
E = {\hbar\over{2e}} J_1 \cos\varphi \; , \; 
J = J_1 \sin\left( \varphi + {{2eAa}\over{\hbar c}}\right)
\label{curj}
\end{equation}

\noindent
where $A$ is the vector potential and a is the distance between islands. 
We relate the Josephson energy $E_J={\hbar\over{2e}} J_1$
to the  magnetization $M$ at low fields, where response is linear.
We use the relation $j=\rho_s A$ between the 
supercurrent and the superfluid density $\rho_s$, and express $M$ via the current.
Assuming for simplicity circular geometry the magnetization is given by

\begin{equation}
M = {1\over{2c}} \int\limits^L_0 2\pi r dr (\vec{j} \times \vec{r})
\end{equation}

\noindent
Taking the integral we relate the magnetization to the magnetic field $H$ and the
superfluid density $\rho_s$

\begin{equation}
M = {1\over{2c}} \rho_s {\pi\over{4}} L^4 H
\label{mag}
\end{equation}

\noindent
We use Eq.\ (\ref{mag}) to deduce the value of $\rho_s$ from the data \cite{efr}; for its
zero temperature value we get $\rho_s(0) = 6.49\times 10^{15}$. 
The superfluid density is a function of reduced temperature, $\rho_s(T)=\rho_s(0)\tau$.
Since we need the superfluid
density at the true melting temperature $T_m$, $\tau$ is defined from the shift in 
melting temperature due to frustration and was measured to be $\tau=0.01$.

Comparing the supercurrent equation with Eq.\ (\ref{curj}) we obtain the Josephson energy
(and the magnitude of current)

\begin{equation}
E = {{\hbar}\over{2e}} J_1 = {{\hbar^2 c}\over{4 e^2}} \rho_s(T_m).
\end{equation}

\noindent
We can now estimate the interaction energy to be $E\approx1.7\times10^4$ K.
Because the melting temperature is of the order of 1 K, its ratio to the interaction energy
$T_c/E = \zeta$ is estimated as $\zeta = 10^{-4}$. 

We now compare Lindemann criterion for thermal and quantum melting.
For the thermal melting considered above we have

\begin{equation}
\langle \rho \rho \rangle \sim T \int {{d^2q}\over{c_{66}q^2}} = a^2_L.
\end{equation}

\noindent
where the integral is over the Brillouin zone. Here we have also substituted $c_{66}q^2$ for the 
actual dispersion law. This rough estimate will be sufficient for our purposes.
Assuming that frustration renormalizes the elastic constant by $c_{66}\sim\kappa\rho_s$,
we get the renormalisation factor $\kappa \sim T/\rho_s a^2_L \approx \zeta /a^2_L$.
Taking for the Lindemann parameter the usual value $a_L\approx 0.1$ we get $\kappa=0.01$.

For quantum melting we conjecture that because the renormalisation $\kappa$ is due to 
frustration (induced by the array) the elastic constant will be renormalized by the same factor.

\begin{equation}
\langle \rho \rho \rangle \sim \int 
\displaystyle{{d^2q d\omega}\over{ {{\omega^2}\over{E_C}} + c_{66}q^2}} 
\approx \displaystyle{\sqrt{ {{E_C}\over{c_{66}}} }} = \tilde{a}^2_L.
\end{equation}

\noindent
Using the renormalisation factor $\kappa$ we get $E_C/\rho_s\sim\kappa\tilde{a}_L^4$.
Thus we expect that the quantum melting happens at kinetic energies which are
at least three orders of magnitude smaller than the interaction energy.

\section{Dissipation effects on vortex motion}

Before we turn to the estimate of array resistance, we have to address the question of energy 
conservation. The quasiclassical calculation of the probability to excite a quasiparticle
considered above assumed implicitly that the energy of the vortex is large enough. Quantitatively
that means that the vortex energy should at least be larger than the quasiparticle gap, otherwise
the vortex would lack the energy to excite the quasiparticle. 

The energy acquired by a vortex driven by the Lorentz force is proportional to the applied 
current so at very low currents it would not be sufficient to excite quasiparticles above
the finite gap. For larger currents the probability to excite a quasiparticles is constant and 
is given by Eq.\ (\ref{kpro}) so at these currents vortex dissipation is linear in its velocity
leading to Ohmic conductivity. Here we estimate smallest currents, $j_0$, at which the 
conductivity remains Ohmic.

There are two effects that make $j_0$ very small. First, a vortex makes many jumps between 
consecutive emission processes and accumulates energy. Second, vortex liquid is incompressible
and retains short range order in a broad range of fields above the melting point, so each
emission process slows down not an individual vortex but a large number of them. This
effect is similar to the M\"ossbauer effect in crystals, but here the momentum can not be 
transfered to the whole number of vortices since there is no long-range order. The effect is
difficult to describe quantitatively, due to the absence of a theory of a strongly correlated
vortex liquid at $T=0$.
We shall attempt therefore only to show that the effect is indeed large.

To see this effect and to estimate the correlation length in the liquid state we perform the following
calculation, similar to the calculation of the Debay-Waller factor. The idea
is that when the momentum is being transfered to the liquid as a whole the quantum state of the
liquid does not change, so it is described by the same wave function after the transfer as before.
The amplitude of such process is 

\begin{equation}
A = \int \Pi dx_i \Psi^*(x_1 ... x_N) 
e^{\displaystyle{\imath {p\over{N}} \sum\limits_i x_i}} \Psi(x_1 ... x_N) \delta(x_1),
\end{equation}

\noindent
where $N$ is the number of vortices, $p$ is the transferred momentum and $\Psi(x_1 ... x_N)$ is
the macroscopic wave function of the vortex liquid. The factor $\delta(x_1)$ singles out 
the coordinate of the island,where the quasiparticle was excited. 
When the vortex interacts with the particular island its coordinate
becomes fixed and therefore we do not need to integrate over it. 

The amplitude $A$ is a function of momentum $p$, vortex number and interaction strength. In a 
true liquid, where no order is present, the wave function $\Psi(x_1 ... x_N)$ depends only on the
relative coordinates of vortices and therefore $A=0$. When vortices are organized in clusters at 
short distances, the correlations decay exponentially like $\exp(-x/\xi)$, where the correlation
length $\xi$ defines cluster size. Therefore for a system of the finite size $L$ the amplitude
$A$ is of the order of unity when $L\sim\xi$, but when the system becomes large, so that $L\gg\xi$,
then the amplitude is exponentially small $A\sim\exp(-L/\xi)$.

Because the vortex-vortex interaction is proportional to the logarithm of distance between
vortices
the vortex liquid can be approximated by the two-dimensional Coulomb gas.
The exact wave-functions of the Coulomb gas are unknown. However there exists a three-body Hamiltonian
with interaction that is Coulomb for long distances, while different (and three-body) for
short distances. The ground-state wave function for this Hamiltonian is known \cite{kan}, and
it was argued that one can use this known wave function to estimate the properties of the 
Coulomb gas. The amplitude $A$ then can be estimated numerically. Due to the limited computer 
availability we performed the calculation for relatively small arrays of up to 20 particles 
in the circular geometry and small interaction parameters
$\alpha=0.5 \div 2.5$ (for comparison, the melting point is at $\alpha=30$ \cite{beda}). But even 
being that far from the melting point we could see the momentum dependence of the amplitude $A$
as described above. Our calculation allows to estimate the correlation length $\xi>5$ in the units
of lattice spacing, which would correspond to the cluster 
size of up to 20-30 vortices and increasing as we increase $\alpha$. 

We now estimate the average energy $E_{cl}$ of a moving cluster due to the external current. 
When the 
current is small the force acting on a vortex is given by $F=J\Phi_0/ac$, where $a$ in the
lattice spacing, $J$ is the current per junction and $\Phi_0$ is the flux quantum. 
The energy acquired by the
vortex after one tunneling jump to the neighboring plaquette is equal to the force times the 
lattice spacing $E=Fa=J\Phi_0/c$. Since the probability to excite a quasiparticle is small 
the excitation is a rare process and the average number of jumps the vortex makes before it excites 
a quasiparticle is inverse probability. Therefore at the moment of quasiparticle 
excitation the vortex would have the energy $E=J\Phi_0/c{\cal P}$. Multiplying this energy 
by the number of particles in the cluster we obtain an estimate of the average energy of the
cluster at the moment of quasiparticle excitation $E_{cl}=N_\xi J\Phi_0/c{\cal P}$. 
 
The probability can be estimated from Eq.\ (\ref{resultd}) using the experimental \cite{exp} values
for $E_C$ and $|\Delta |$. We get ${\cal P}\approx 0.001$. Therefore the cluster energy can be 
estimated as ${E_{cl}=N_\xi J \cdot 1\times 10^{2}}$ K, where the current is measured in nanoampers. 

The system has linear response when the average cluster energy is larger than the gap 
$|\Delta | \approx 2$ K.
For very small currents, when $E_C < |\Delta |$ the system would exhibit non-linear 
current-voltage curves. For a cluster size $N_\xi=100$ the current value where this non-linearity
would be observable can be estimated by setting $E_{cl}$ equal to the gap and using the above
estimate for the cluster energy. We get $J_0 = 0.1\times 10^{-3}$ nA. The currents used in 
experiment are of the order of $J\approx 0.1$ nA. Therefore it is likely that
in the experimentally observable case the array is in the (pseudo) linear regime.
   
Thus the coupling to quasiparticles results in dissipation described by
linear response in contrast to the dissipation due to coupling to the
acoustic phase modes (spin waves) \cite{lark,ecke}. The effective
action obtained in Ref. \cite{lark} has a dissipation term which is
proportional to $\omega^2 \ln\omega$. For the slow vortex motion we
consider (due to very small currents discussed above) this term is small
compared to the linear term Eq. (35) which is determined by the energy
scale $E_J$ rather than the frequency. For larger currents the situation is
different and the two effects become comparable.

For the (pseudo) linear response regime we now consider how the vortex motion is affected by the 
dissipation. Our goal is to obtain an expression for the total array resistance which 
arises due to dissipation Eq.\ (\ref{resultd}), therefore we now consider a macroscopic equation
of vortex motion, averaged over the whole array following Ref. \cite{vdz}. Note that after
averaging this equation (Eq. \ (\ref{eqm})) does not describe the microscopic coupling between
vortices and quasiparticles, which in general is non-linear.

Under the influence of the driving current $J$ a vortex is moving in a direction perpendicular to
the current flow:

\begin{equation}
\Gamma_0^{-1} \ddot{x} + {\cal W} 
\Gamma_0^{-1} \dot{x} = {{{\it \Phi}_0 J a}\over{c}},
\label{eqm}
\end{equation}

\noindent
where $x$ is the vortex position ${\it \Phi}_0$ is the flux quantum, 
$\Gamma_0$ is the tunneling rate, defined in Eq.\ (\ref{trate}),
which provides the measure for the vortex mass and $a$ is the lattice spacing. 
The lattice potential was
taken into account when we calculated the dissipation and the tunneling rate, therefore it does not
appear in Eq.\ (\ref{eqm}).

If the driving current is constant then the vortices move with constant velocity. If in the Cartesian
coordinate system the current flows along $y$ axis, then vortices move along $x$ (magnetic field is 
along $z$ axis, perpendicular to the $xy$ plane of the array) and their velocity is

\begin{equation}
v_x = {{\pi\hbar J a}\over{e {\cal W}}} \Gamma_0 .
\end{equation}

\noindent
The potential difference caused by the time dependent phase is

\begin{equation}
U = {{\hbar}\over{2e}} {{\partial\phi}\over{\partial t}}.
\end{equation}

\noindent
When one vortex is moving across the array (in time $t=d/v_x$, where $d$ is the size of the
array) the phase changes by $2\pi$. To obtain the total voltage the effect of one vortex
should be multiplied by their number $n_v = B d^2 / {\it \Phi}_0$

\begin{equation}
U = \pi^2 {\hbar \over{e^2}} \; {{(\hbar\Gamma_0)}\over{{\cal W}}} 
\; {{Bda}\over{{\it \Phi}_0}} J
\end{equation}

\noindent
The current per junction $J$ can be obtained from the total current $I$ as $J=Ia/d$
(assuming a square array). 
The coefficient of proportionality between the voltage $U$ and the current $I$ 
is the array resistance 

\begin{equation}
R = \pi^2 {\hbar \over{e^2}} \; {{(\hbar\Gamma_0)}\over{{\cal W}}} \; {{B a^2}\over{{\it \Phi}_0}}.
\label{res}
\end{equation}

\section{Conclusions}

We have considered the vortex tunneling in Josephson-junction arrays at zero temperature. Using the simple
quantum-mechanical analogy we have showed that such tunneling is accompanied by small dissipation due
to quasiparticle excitations in the superconducting islands. Even in the presence of the large
quasiparticle gap the probability of such excitations is found to contain only power-law
smallness in the (small) ratio of the characteristic tunneling frequency to the gap.

Our main result is the resistance of the vortex liquid Eq.\ (\ref{res}), which is due to this 
dissipation. The result is valid for not too small driving currents (for experimental setup 
\cite{exp} we estimate $J>J_0=0.1\times 10^{-3}$ nA), 
for which the system is in the (pseudo) linear response regime. 
Our results provide the quantum-mechanical mechanism of dissipation in Josephson-junction arrays. The 
resistance Eq.\ (\ref{res}) is independent of temperature (due to it's quantum-mechanical origin) 
in agreement with the experiment \cite{exp}.

Our argument is applicable to the vortex liquid just above the melting point, where
the vortex liquid retains short-range crystalline order. Then the vortices are 
strongly interacting and the external momentum can be transfered to a large number of vortices. This
is similar to the M\"ossbauer effect in crystals where the external momentum is transferred to the whole 
crystal. In the vortex liquid such a transfer is impossible due to the absence of long-range order,
but the transfer to a finite size cluster remains possible. Such a cluster involves a large number
of vortices, therefore its energy is much more than the gap in the quasiparticle spectrum, 
allowing excitations above the gap. Estimating the minimal cluster size from our numerical data
we got that in experimental conditions the energy contained in such cluster is always larger than the
gap; it would get smaller than the gap only for very small currents $J<J_0$.

However these estimates depend crucially on the structure of the strongly correlated vortex liquid
formed when the vortex lattice melts. We have argued that the problem is exacerbated by the
frustration imposed on the vortex lattice by the underlying array structure. The frustration
reduces even further the kinetic energy needed for the melting.
The self-consistent theoretical description of the normal liquid of vortices, however, remains to
be an unresolved question. In particular it is not clear whether the existence of the normal liquid
is due to the dissipation effects or it is in fact possible to form a normal liquid in the absence
of dissipation.

One of the possible descriptions of the strongly correlated vortex liquid is a dilute gas of
dislocations in the vortex crystal similar to the hexatic phase appearing in 2D thermal melting\cite{halp}.
In this phase the vortex flow is due to the motion of dislocations. The motion of each 
single dislocation transfers the whole row of vortices across the system. 
Here the number of moving vortices scales with the system size so in the thermodynamic limit
the combined energy of these vortices becomes infinite and linear response persists to zero currents.

The detailed description of the strongly correlated vortex liquid is a subject of future work.

\end{document}